\documentclass[twocolumn,usenatbib,floatfix]{aastex62}
\usepackage{graphicx}
\usepackage{float}

\graphicspath{{./}{figures/}}

\received{January 1, 2018}
\revised{January 7, 2018}
\accepted{\today}
\submitjournal{ApJ}

\shorttitle{}
\shortauthors{A. Bonafede et al.}

\begin{document}

\title{The Coma cluster at LOFAR frequencies I:  insights into particle acceleration mechanisms in the radio bridge.}

\correspondingauthor{Annalisa Bonafede et al}
\email{annalisa.bonafede@unibo.it}

\author{A. Bonafede}
\affiliation{DIFA - Universit\'a di Bologna, via Gobetti 93/2, I-40129 Bologna, Italy }
\affiliation{INAF - IRA, Via Gobetti 101, I-40129 Bologna, Italy; IRA - INAF, via P. Gobetti 101, I-40129 Bologna, Italy;} 
\affiliation{Universit\"at Hamburg, Hamburger Sternwarte, Gojenbergsweg 112, 21029, Hamburg, Germany; }

\author{G. Brunetti}
\affiliation{INAF - IRA, Via Gobetti 101, I-40129 Bologna, Italy; IRA - INAF, via P. Gobetti 101, I-40129 Bologna, Italy;}

\author{F. Vazza}
\affiliation{DIFA - Universit\'a di Bologna, via Gobetti 93/2, I-40129 Bologna, Italy }
\affiliation{INAF - IRA, Via Gobetti 101, I-40129 Bologna, Italy; IRA - INAF, via P. Gobetti 101, I-40129 Bologna, Italy;}
\affiliation{Universit\"at Hamburg, Hamburger Sternwarte, Gojenbergsweg 112, 21029, Hamburg, Germany; }

\author{A. Simionescu}
\affiliation{SRON Netherlands Institute for Space Research, Sorbonnelaan 2, 3584 CA Utrecht, The Netherlands;}
\affiliation{Leiden Observatory, Leiden University, PO Box 9513, 2300 RA Leiden, The Netherlands;}
\affiliation{Kavli Institute for the Physics and Mathematics of the Universe (WPI), The University of Tokyo, Kashiwa, Chiba 277-8583, Japan; }

\author{G. Giovannini}
\affiliation{DIFA - Universit\'a di Bologna, via Gobetti 93/2, I-40129 Bologna, Italy }
\affiliation{INAF - IRA, Via Gobetti 101, I-40129 Bologna, Italy; IRA - INAF, via P. Gobetti 101, I-40129 Bologna, Italy;} 

\author{E. Bonnassieux}
\affiliation{DIFA - Universit\'a di Bologna, via Gobetti 93/2, I-40129 Bologna, Italy }
\affiliation{INAF - IRA, Via Gobetti 101, I-40129 Bologna, Italy; IRA - INAF, via P. Gobetti 101, I-40129 Bologna, Italy;}

\author{T. W. Shimwell}
\affiliation{ASTRON, Netherlands Institute for Radio Astronomy, Oude Hoogeveensedijk 4, 7991 PD, Dwingeloo, The Netherlands;}
\affiliation{Leiden Observatory, Leiden University, PO Box 9513, 2300 RA Leiden, The Netherlands;}

\author{M. Br\"uggen}
\affiliation{Universit\"at Hamburg, Hamburger Sternwarte, Gojenbergsweg 112, 21029, Hamburg, Germany; }

\author {R. J. van Weeren}
\affiliation{Leiden Observatory, Leiden University, PO Box 9513, 2300 RA Leiden, The Netherlands;}

\author{A. Botteon}
\affiliation{Leiden Observatory, Leiden University, PO Box 9513, 2300 RA Leiden, The Netherlands;}

\author{M. Brienza}
\affiliation{DIFA - Universit\'a di Bologna, via Gobetti 93/2, I-40129 Bologna, Italy }
\affiliation{INAF - IRA, Via Gobetti 101, I-40129 Bologna, Italy; IRA - INAF, via P. Gobetti 101, I-40129 Bologna, Italy;}

\author{R. Cassano}
\affiliation{INAF - IRA, Via Gobetti 101, I-40129 Bologna, Italy; IRA - INAF, via P. Gobetti 101, I-40129 Bologna, Italy;}

\author{A. Drabent}
\affiliation{Th\"uringer Landessternwarte, Sternwarte 5, D-07778 Tautenburg, Germany }

\author{L. Feretti}
\affiliation{INAF - IRA, Via Gobetti 101, I-40129 Bologna, Italy; IRA - INAF, via P. Gobetti 101, I-40129 Bologna, Italy;}

\author{F. de Gasperin}
\affiliation{Universit\"at Hamburg, Hamburger Sternwarte, Gojenbergsweg 112, 21029, Hamburg, Germany; }

\author{F. Gastaldello}
\affiliation{INAF - IASF Milano, via A. Corti 12, 20133 Milano, Italy;}

\author{G. di Gennaro}
\affiliation{Leiden Observatory, Leiden University, PO Box 9513, 2300 RA Leiden, The Netherlands;}

\author{M. Rossetti}
\affiliation{INAF - IASF Milano, via A. Corti 12, 20133 Milano, Italy;}

\author{H. J. A. Rottgering}
\affiliation{Leiden Observatory, Leiden University, PO Box 9513, 2300 RA Leiden, The Netherlands;}

\author{C. Stuardi}
\affiliation{DIFA - Universit\'a di Bologna, via Gobetti 93/2, I-40129 Bologna, Italy }
\affiliation{INAF - IRA, Via Gobetti 101, I-40129 Bologna, Italy; IRA - INAF, via P. Gobetti 101, I-40129 Bologna, Italy;}

\author{T. Venturi}
\affiliation{INAF - IRA, Via Gobetti 101, I-40129 Bologna, Italy; IRA - INAF, via P. Gobetti 101, I-40129 Bologna, Italy;}

\nocollaboration

\begin{abstract}
Radio synchrotron emission from the bridges of low-density gas connecting galaxy clusters and groups is a challenge for particle acceleration processes. In this work, we analyse the Coma radio bridge using new LOw Frequency ARray (LOFAR) observations at 144 MHz. LOFAR detects the bridge and its substructures with unprecedented sensitivity and resolution. We find that the radio emission peaks on the NGC~4839 group. Towards the halo, in front of the NGC~4839 group, the radio brightness decreases and streams of radio emission connect the NGC~4839 group to the radio relic. Using X-ray observations, we 
find that thermal and non-thermal plasma are moderately correlated with a sub-linear scaling. We use archival radio data at 326 MHz to constrain the spectral index in the bridge, and quantify the distribution of particles and magnetic field at different frequencies. We find that the spectrum is steeper than $-1.4 \pm 0.2$, and that the emission could be clumpier at 326 MHz than at 144 MHz. Using cosmological simulations and a simplified approach to compute particle acceleration, we derive under which conditions  turbulent acceleration of mildly relativistic electrons could generate the radio emission in the bridge. Assuming that the initial energy ratio of the seed electrons is $3 \cdot 10^{-4}$ with respect to the thermal gas, we are able to reproduce the observed luminosity. Our results suggest that the seed electrons released by radiogalaxies in the bridge and the turbulence generated by the motion of gas and galaxies are essential to produce the radio emission.

\end{abstract}



\section{Introduction}
\label{sec:intro}

Galaxy clusters accrete matter through filaments of the so-called cosmic web. Mergers between galaxy clusters and groups are extremely energetic, with up to $10^{64}$ ergs released in the intracluster medium (ICM). A fraction of this energy could be channeled into magnetic field amplification and particle acceleration leading to  synchrotron radio emission.  \\
Different types of radio sources are associated with the ICM: mini halos, radio halos, and radio relics, are sources located in the cluster core, in the cluster central region, and at the cluster periphery, respectively. They all have a low surface brightness ($\sim 1 \mu \rm{Jy/arcsec^2}$) at GHz frequencies, and a steep radio spectrum, with a spectral index $\alpha < -1$\footnote{Throughout this paper, we define the spectral index $\alpha$ as $S(\nu) \propto \nu^{\alpha}$, where S in the flux density at the frequency $\nu$} that make them brighter at low radio frequencies. We refer the reader to the reviews by \citet{BJ14} and \citet{vanWeeren19} for further information on these sources.\\
The Coma cluster hosts the most famous and first detected radio halo  \citep{Large59} and radio relic \citep{Ballarati81,Giovannini91}. A bridge of low surface brightness connecting the two was first reported by \citet{kim89} and further studied afterwards (e.g. \citealt{Venturi90}, \citealt{BrownRudnick11}). For many years, this bridge has been a unique case. Recently, the advent of low frequency ($\nu< 1 ~\rm{GHz}$) instruments has lead to the discovery of bridges of radio emission connecting 
clusters and merging groups \citep{Bonafede18,Botteon19}. 
More recently, large-scale bridges of radio emission have also been detected between massive clusters in a pre-merger state \citep{Govoni19,Botteon20bridge}.
The detection of diffuse and low-brightness radio emission is possible thanks to 
the new facilities, such as the LOw Frequency ARray \citep[LOFAR][]{vanHaarlem}, that permit to detect large-scale radio emission while keeping high angular resolution to disentangle the  embedded sources.

\subsection{Large-scale bridges and the Coma bridge}

The discovery of large-scale radio bridges connecting A399 and A401 \citep{Govoni19} calls for the existence of different (re)acceleration scenarios than those used to explain radio halos and relics. Recently, \cite{BV20} have proposed that stochastic acceleration of relativistic electrons by turbulence could explain the observed emission. According to this model,
a key role is played by the complex dynamics along the filament connecting the two clusters.
Indeed, cosmological simulations indicate that the chaotic accretion of matter along the filament could generate turbulence. The resulting radio emission is expected to be volume-filling and with a steep spectrum ($\alpha <$ -1.3).
\\
The Coma radio bridge has a different spatial scale and is located in a different environment than the large-scale bridge found between A399 and A401.
However, several optical, millimetric, and  X-ray, studies indicate that the bridge is located in a very active dynamical region. Hence, we will investigate in this work if turbulence could also play a role in the post-merging Coma bridge.
  
 \subsection{The dynamical status in the Coma-bridge region} 
\citet{Malavasi20} have recently analysed the large-scale structure around the Coma cluster. They have found that Coma is located at the centre of a network of three filaments, one of which connects Coma to the cluster A1367, a massive ($M_{\rm SZ}=1.7 \pm 0.1 \times 10^{14}$ \citealt{Planck13}),  dynamically young cluster, that is undergoing a merger between its two sub-groups \citep{Ge19}.\\ The group  NGC~4839 and the radio-galaxy NGC~4789 are located at the periphery of Coma, towards A1367, and close to the filament of galaxies and groups that connects the two clusters. 
\citet{Lyskova19} have analysed the
interaction between the NGC~4839 group and the Coma cluster, using X-ray observations from XMM-Newton and Chandra and comparing their analysis with numerical simulations. 
They observe a tail of cold gas to the south-west of the group, and a ``sheath'' region of hotter gas that envelops the group.
They conclude that the most likely scenario is the one where the group is at its second infall towards the Coma cluster after passing the apocenter.
The shock that is responsible for the relic would be due to the first infall of the group.\\
More recently, \cite{Mirakhor20} have analysed XMM-Newton and  Planck Sunyaev Zel'dovich data of the Coma cluster. They found a significant entropy deficit in SW sector of Coma, towards A1367. This entropy deficit is consistent with theoretical expectations for filamentary gas streams from the cosmic web towards the Coma cluster.

Overall, several independent works indicate that the SW region of Coma, where the radio bridge is located, is a dynamically active region where the NGC~4839 group is falling into Coma at its second passage, an out-moving  shock wave is likely originating the relic, and filaments of gas are accreting 
onto the main cluster along the filament that connects Coma with A1367.
All these motions could inject turbulence in the bridge region and possibly accelerating a seed population of mildly relativistic electrons to originate the radio bridge.


In this work, we use LOFAR and Westerbork Synthesis Radio Telescope (WSRT) observations to analyse the properties of the Coma bridge. We also use X-ray observations with the ROentgen SATellite (ROSAT) to understand the connection between thermal and non-thermal plasma in the bridge.
A detailed study of the halo and relic will be the subject of a forthcoming paper.\\ The paper is structured as follows: In Sec. \ref{sec:obs} we describe the observations and data reduction techniques, and in Sec. \ref{sec:bridge} we analyse the bridge properties and its connections with the cluster thermal plasma. 
In Sec. \ref{sec:sim}, we use the cosmological simulations of a Coma-like cluster to inspect under which conditions the Coma bridge can be powered by turbulence.
Finally, we conclude in Sec. \ref{sec:conclusions}.
Throughout this paper, we assume a $\Lambda$CDM cosmological model, with $\rm{H_0=69.6 km/s/Mpc}$, $\Omega_M=0.286$, $\Omega_{\Lambda}=0.714$. At the Coma redshift (z=0.0231) the angular to linear scale is 0.47$\arcsec$/kpc.

\section{Data reduction and imaging}
\label{sec:obs}
\subsection{LOFAR observations}
The data used in this work are part of the LOFAR Two Meter Sky Survey (LoTSS, see \citealt{Shimwell19}). They consist of two pointings of 8 hours each that have been calibrated independently following a slightly different procedure than the standard one described in \cite{Shimwell19} and Shimwell et al (in prep). Specifically, for the pointing P192+27, a larger region of calibration and imaging has been used, to  account for a strong off-axis source located outside the beam first side lobe. For both P195+27 and P192+27 pointings, we have kept all baselines for calibration, while the LoTSS images and calibration is done with an inner cut of the recorded visibilities below 0.1 km. This cut allows one to easily eliminate radio frequency interferences on the short baselines, and filters out extended emission (larger than 34 arcminutes in the sky.), e.g. coming from our Galaxy, that is hard to deconvolve and hence would make the calibration difficult.
Hence, it cannot be applied to the Coma cluster, that extends for more than 2$^{\circ}$.\\
In order to image the diffuse emission of the Coma cluster, we have performed the following steps:
\begin{enumerate}
\item Subtraction of sources outside the Coma cluster region: sources outside the inner 1.5$^{\circ}$ circle radius have been subtracted using KillMS and DDF \citep{Smirnov15} in order to speed up the following imaging steps. Specifically, after imaging 
the model components have been corrupted by the antenna gains and subtracted by the UV visibilities. This procedure has been repeated twice with different inner UV-range (baselines longer than 0.1 km, and all baselines) in order to remove both compact and diffuse emission outside the Coma cluster. The resulting dataset consists of emission from the Coma cluster only.
\item The dataset has been reimaged using an inner UV-cut of 0.3 km, in order to  
obtain a model for point-like and extended sources inside the Coma cluster that we have subtracted from the data. The model components corresponding to NGC 48im39 and NGC 4789 have been excluded from the subtraction, as the emission from those tails merges with the diffuse emission of the bridge and of the relic. The resulting datasets consist of only diffuse emission in the Coma cluster region. Some residual emission from embedded sources could still be present, but its contribution should be negligible.
\end{enumerate}


 Imaging has been performed using DDF \citep{Tasse18}, to apply the directional gains derived during calibration and to take into account the differential beam pattern throughout the Coma field.
 We used the Sub Space Deconvolution (SSD) algorithm that is generally robust in the deconvolution of extended emission (\citealt{Tasse18} and ref. therein). However, the emission is the Coma field is very extended and characterised by a low-surface brightness. Even with SSD, multiple cycles of deconvolution with optimised masks, each covering a different portion of the halo/bridge have been performed. Although this procedure is not optimal, it gives better results than the HMP (Hybrid  Matching  Pursuit) algorithm, and no other algorithm is implemented in DDF at the time of writing.
 We have checked that the residual images did not present regions of un-deconvolved emission in the region of interest.\\
The  images made from observations of the pointing P192+27 have a higher noise with respect to those from P195+27 (see Tab. \ref{tab:obs}). From a comparison between the images presented in this work and those of the Data Release 2 (Shimwell et al. in prep) we conclude that this is due to interference and/or flags occurring in the short baselines.

\begin{figure*}
\includegraphics[width=1.1\textwidth]{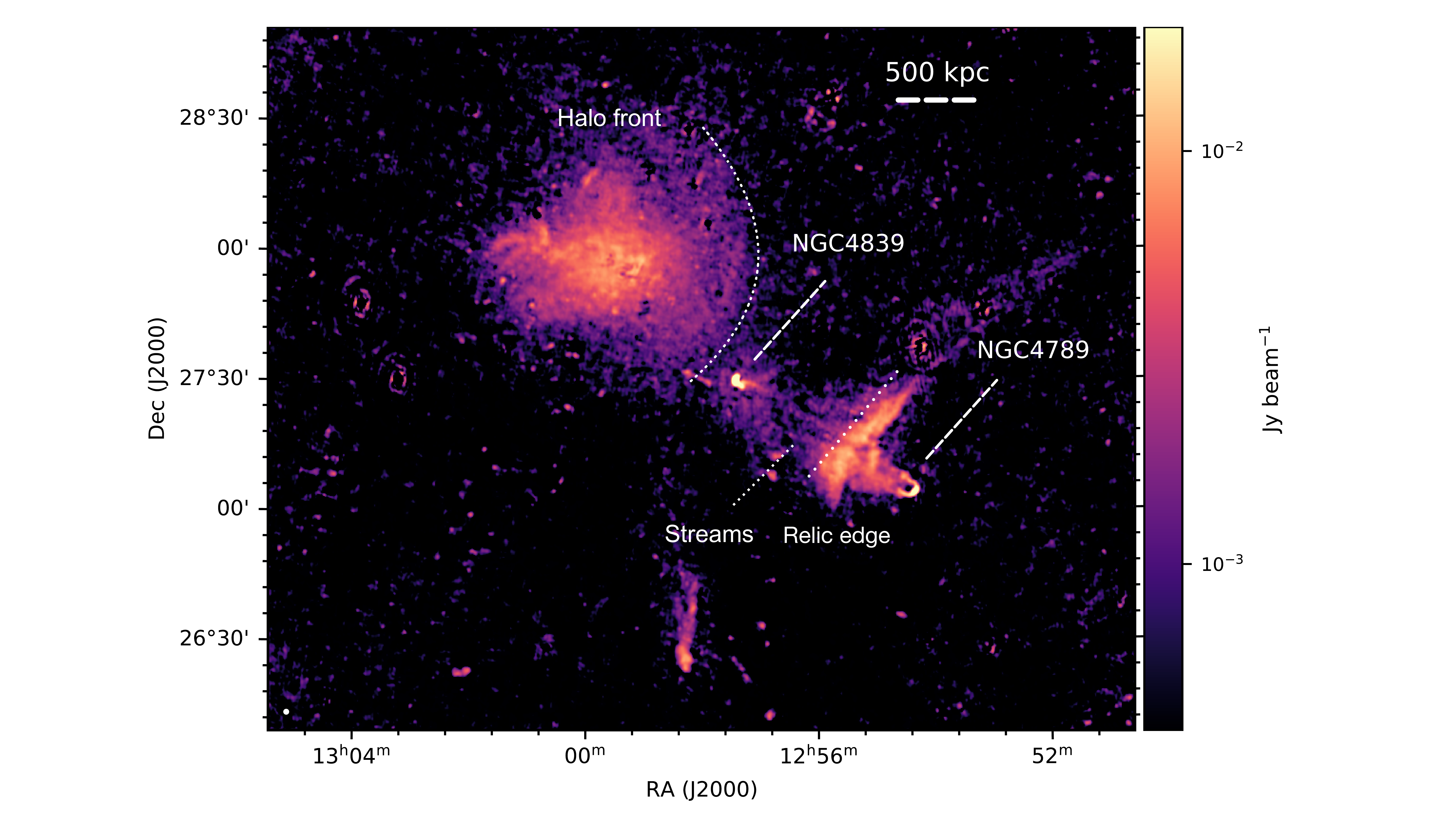}
\caption{LOFAR emission of the Coma field. The restoring beam is $\sim 45\arcsec \times 45\arcsec$, and is shown in the bottom left corner of the image. The noise is 0.25 mJy/beam. Sources embedded in the diffuse emission have been subtracted, except for the two tails NGC~4839 and NGC 4789. Relevant sources and features are labelled.}
\label{fig:Coma_extended}
\end{figure*}

The images of the two pointings have been restored to the same beam, corrected for the primary beam response, and finally combined by weighting with the inverse of their noise.
The combined image of the Coma cluster diffuse emission (plus the two tails) is shown in Fig.~\ref{fig:Coma_extended}.
The absolute flux scale has been checked to ensure consistency with the LOFAR DR2 images (Shimwell et al, in prep).
The error on the LOFAR absolute flux scale is assumed to be 15\%. Details on the observations are listed in Table \ref{tab:obs}.

\subsection{X-ray observations}
We have analyzed the data from four ROSAT Position Sensitive Proportional Counter (PSPC) observations of the Coma cluster extending out to radii of approximately 60--70$^\prime$. These observations, with identifiers rp800006N00, rp800013N00, rp800005N00, and rp800009N00, were carried out between 1991 June 16--18 and their combined clean exposure time is 78 ks. The data were reduced using the Extended Source Analysis Software \citep{snowden1994}. Background and exposure-corrected images in three energy bands (0.7--0.9, 0.9--1.3, and 1.3--2.0 keV, respectively) were combined, after removing artifacts associated with the detector edges.\\
In order to account for the Galactic foreground, 
 whose intensity may be non-negligible in the low surface brightness bridge, we fitted a $\beta-$model \citep{betamodel} plus a constant to the ROSAT mosaic. We excluded from the fit the central 3$^\prime$ region of the Coma ICM, the contaminating point-like sources, and the region contained between the azimuths  -78--2 degrees from the $W$ axis which correspond to the bridge region and the NGC~4789 sub-halo. 
The best-fit constant thus determined has been subtracted from the exposure-corrected image to eliminate the Galactic foreground and emission from unresolved distant AGN.

\begin{table}
\caption{Observation details}
\begin{tabular}{lccc}
\hline\hline
LOFAR pointing & Time & $\theta_{rest}$  &  $\sigma_{rms}$ \\
                &  [h]          &               &    mJy/beam\\
LOFAR P192+27 &  8  &  30\arcsec $\times$ 18\arcsec   &  0.35 \\
LOFAR P195+27 &  8  &   30\arcsec $\times$ 25\arcsec  & 0.25 \\
\hline\hline
\multicolumn{4}{l}{Col 1: LoTSS Pointing name; Col 2: observing time;}\\
\multicolumn{4}{l}{Col 3: Restoring beam; Col 4: rms noise of the image.}
\end{tabular}	
\label{tab:obs}
\end{table}

\begin{table}
\caption{Image details}
\begin{tabular}{lccc}
\hline\hline
Freq & $\theta_{rest}$ & $\sigma_{rms}$ & Fig. \\
MHz    &  & mJy/beam & \\
144   & 30\arcsec $\times$ 30\arcsec &  0.2 & \ref{fig:Coma_extended}, \ref{fig:2res}\\      
144   & 45\arcsec $\times$ 45\arcsec &  0.25 & \ref{fig:2res}, \ref{fig:corr}\\ 
144   & 150\arcsec $\times$ 100\arcsec &  1.0 & \\ 
326   & 150\arcsec $\times$ 100\arcsec &  0.5 &  \ref{fig:spix_profile} \\
\hline\hline
\multicolumn{4}{l}{Col 1: Observing frequency; Col 2: Restoring beam; }\\
\multicolumn{4}{l}{Col 3: rms noise of the image; Col 4: Figure of merit.}
\end{tabular}\\	
\label{tab:image}
\end{table}

\begin{table}
\caption{Bridge properties}
\begin{tabular}{lcccc}
\hline\hline
Freq              &  LAS & LLS & Flux density & P  \\ 
MHz   &$^{\prime}$  & kpc & Jy & ergs/s/Hz \\
326            & 30 & 840 & 0.47$\pm$0.05 &5.8 $\pm$ 0.6 $10^{30}$ \\
144            & 34 & 940 & 1.6$\pm$ 0.2  & 2.0$\pm$0.2 $10^{31} $ \\  
\hline\hline

\multicolumn{5}{l}{Col 1: Frequency of the observation; Col 2: Largest angular}\\
\multicolumn{5}{l}{  scale of the bridge; Col 3: Largest linear scale using the  }\\
\multicolumn{5}{l}{cosmological model adopted in this paper. Col 4: Flux  }\\
\multicolumn{5}{l}{ density of the bridge; Col 5: Corresponding }\\
\multicolumn{5}{l}{  monochromatic power; .}
\end{tabular}
\label{tab:bridge}
\end{table}

\begin{figure*}
\includegraphics[width=\textwidth]{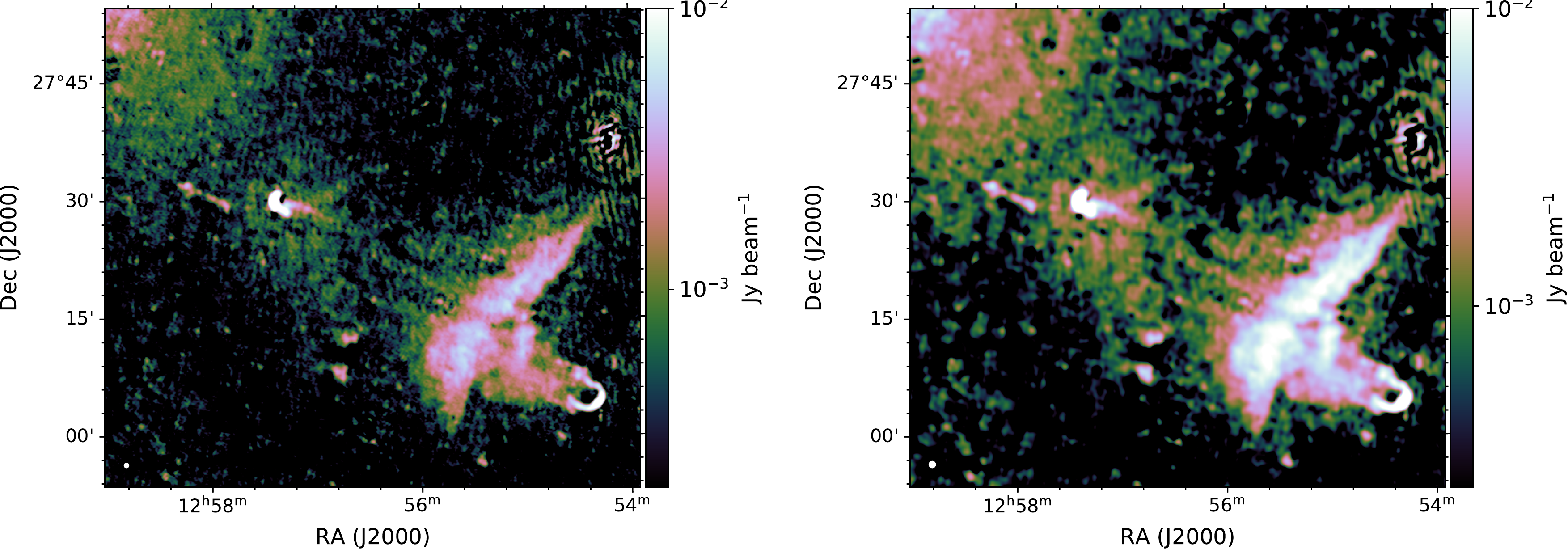}
\caption{Left panel: LOFAR emission from the Coma bridge at 30\arcsec $\times$ 30\arcsec. The emission below 3$\sigma_{\rm rms}$ (0.6 mJy/beam) is displayed in black and shows a brightess decrement between the halo and the bridge. Right panel: LOFAR emission from the Coma bridge at 45\arcsec $\times$ 45\arcsec. The emission below 2 $\sigma_{\rm rms}$, corresponding to 0.5 mJy/beam, is displayed in black. The bridge appears connected to the halo.}
\label{fig:2res}
\end{figure*}

\begin{figure}
\includegraphics[width=\columnwidth]{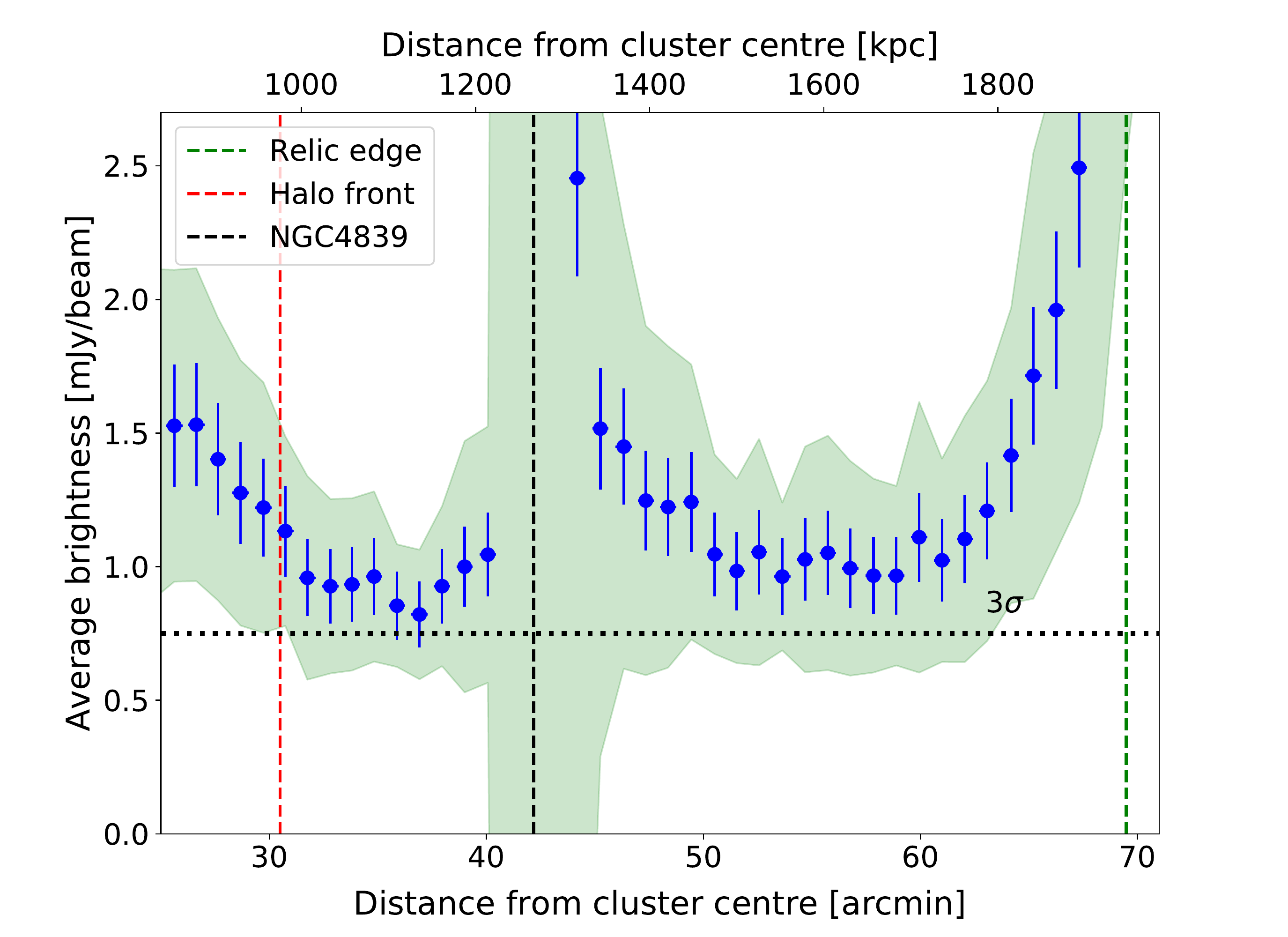}
\caption{Brightness profile along the bridge computed in boxes parallel to the relic main axes. The green area indicates the scatter inside each box. The positions of the relic edge, halo front, and of the NGC~4839 group are marked by dashed lines. The brightness is computed from the image at 45\arcsec resolution.}
\label{fig:profile}
\end{figure}

\section{The Coma radio bridge} 
\label{sec:bridge}
The LOFAR image confirms the presence of a bridge of low surface brightness located in-between the cluster and the radio relic (Fig.~\ref{fig:Coma_extended}).\\
Thanks to the sensitivity of LOFAR, the bridge is detected with a high significance, and its substructures are mapped in detail.
The higher resolution of the LOFAR data, with respect to previous studies, shows that the radio emission does not uniformly cover the whole region between the cluster and the relic. From Fig.~\ref{fig:Coma_extended}, we see that the emission surrounds the  NGC~4839 group and that the group and the relic are connected through streams of radio emission, similar to those observed in the ``Toothbrush" radio relic, in the shock downstream region \citep{Kamlesh20}.
 These streams could mark regions of enhanced magnetic field or non-thermal plasma filaments
resulting from the motion of the source NGC~4789 towards South-West.  Understanding their origin goes beyond the scope of this paper. We note that polarisation images as well as spectral index images at high resolution would give important information about the origin of the streams. Indeed, they would allow us to infer the orientation of the magnetic field and constrain the dynamics of relativistic particles. 
\\
At resolution of 30\arcsec, the emission from the bridge above 3$\sigma_{\rm rms}$ appears disconnected from the halo. At the same time, though, the bridge emission is better recovered with a 45\arcsec \ resolution.
 Indeed, the flux density measured from the image at 45\arcsec  is higher than that measured from the image at 30\arcsec. This indicates that we are genuinely recovering more diffuse emission at lower resolution, and that we are not just convolving with a larger restoring beam the emission recovered at 30 \arcsec.
At 45\arcsec, the bridge appears connected to the radio halo (Fig.~\ref{fig:2res}).
We will use the 45\arcsec\, image in the following analysis, deeper observations would be needed to exclude that the connection observed at 45 \arcsec is due to the convolution of the emission with the lower resolution restoring beam.\\
The radio bridge has a Largest Linear Size LLS$\sim 940$ kpc and a flux density $S=1.6\pm 0.2$ Jy. Other details are listed in Table \ref{tab:bridge}.
In Fig.~\ref{fig:profile}, we show the radio brightness profile along the bridge. The brightness is averaged inside rectangular boxes having the main axis perpendicular to the bridge main axis, i.e. parallel to the relic's main axis. The boxes are $\sim 600 \times 50 $ kpc, and only the emission above 2 $\sigma_{rms}$ has been integrated within each box.
The average brightness of the bridge is a factor $\sim 6-8$ lower than in the main body of the relic. It declines gradually from the relic inner edge towards the NGC~4839 group, and it remains fairly constant throughout the bridge, except for an increase and a higher dispersion  observed in the region of the NGC~4839 group.\\
In front of the NGC~4839 group, towards the Coma cluster, the bridge appears to merge with the halo, and it is not obvious to separate the emission of the two.
However, the western part of the halo shows a semi-circular region that could be linked to the shock front detected in the X-rays \citep{SuzakuComa} and by Planck \citep{PlanckComa}. The northern part of this feature has been already identified by \cite{BrownRudnick11} and called ``halo-front". LOFAR observations show that the front extends to the SW of the cluster. A detailed analysis of this front will be presented in a forthcoming paper. Here, we refer to the radio halo as the emission that is confined by the front (see Fig.\ref{fig:Coma_extended}), and to the radio bridge as the emission that is beyond the front with respect to the Coma cluster centre.\\
Assuming that radio emitting electrons are (re)accelerated by the shock at the position of the relic, we can estimate the cooling region of the particles, i.e. the part of the bridge that can be explained without assuming any further acceleration mechanism. With a mean magnetic field in the bridge region $\langle B \rangle =$ 0.5 --  2$\mu \rm{G}$ \citep{va18mhd,Bonafede13} the cooling length behind a shock re-accelerating a fossil population of relativistic electrons would be $\sim$150 - 200 kpc at 144 MHz \citep[e.g.][]{KangRyu12}. Since the radio emission is observed on scales that are a factor 4 higher than the electron cooling length, we can conclude that an additional acceleration mechanism is operating in the bridge.
\begin{table}
\centering
\caption{Spectral properties of the bridge}
\begin{tabular}{lcc}
\hline\hline
$\alpha_{patches}$ & $\alpha_{L}$ & $\alpha_{U}$ \\
 -1.3 $\pm$ 0.2 &  $-1.6 \pm 0.2$& $-1.4 \pm 0.2$\\
\hline\hline
\multicolumn{3}{l}{Col 1: Spectral index of the emission detected with }\\
\multicolumn{3}{l}{WSRT above 2$\sigma_{\rm rms}$;}\\
\multicolumn{3}{l}{Col 2, 3: lower, upper limit }\\
\multicolumn{3}{l}{ to the spectral index in the bridge region.}
\end{tabular}
\label{tab:alpha}
\end{table}

\subsection{Spectral index analysis}
\begin{figure*}
\includegraphics[width=1.1\columnwidth]{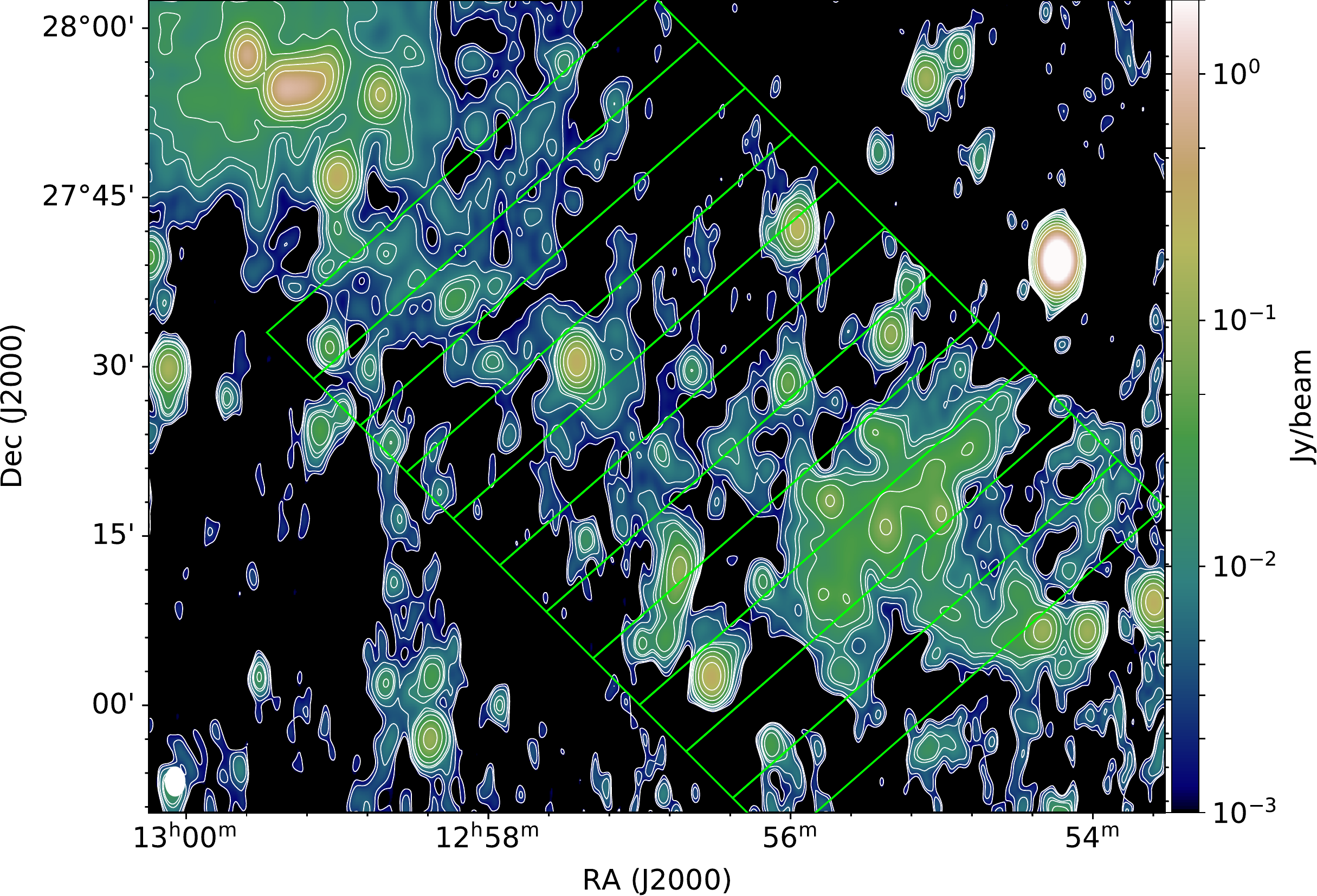}
\includegraphics[width=1.1\columnwidth]{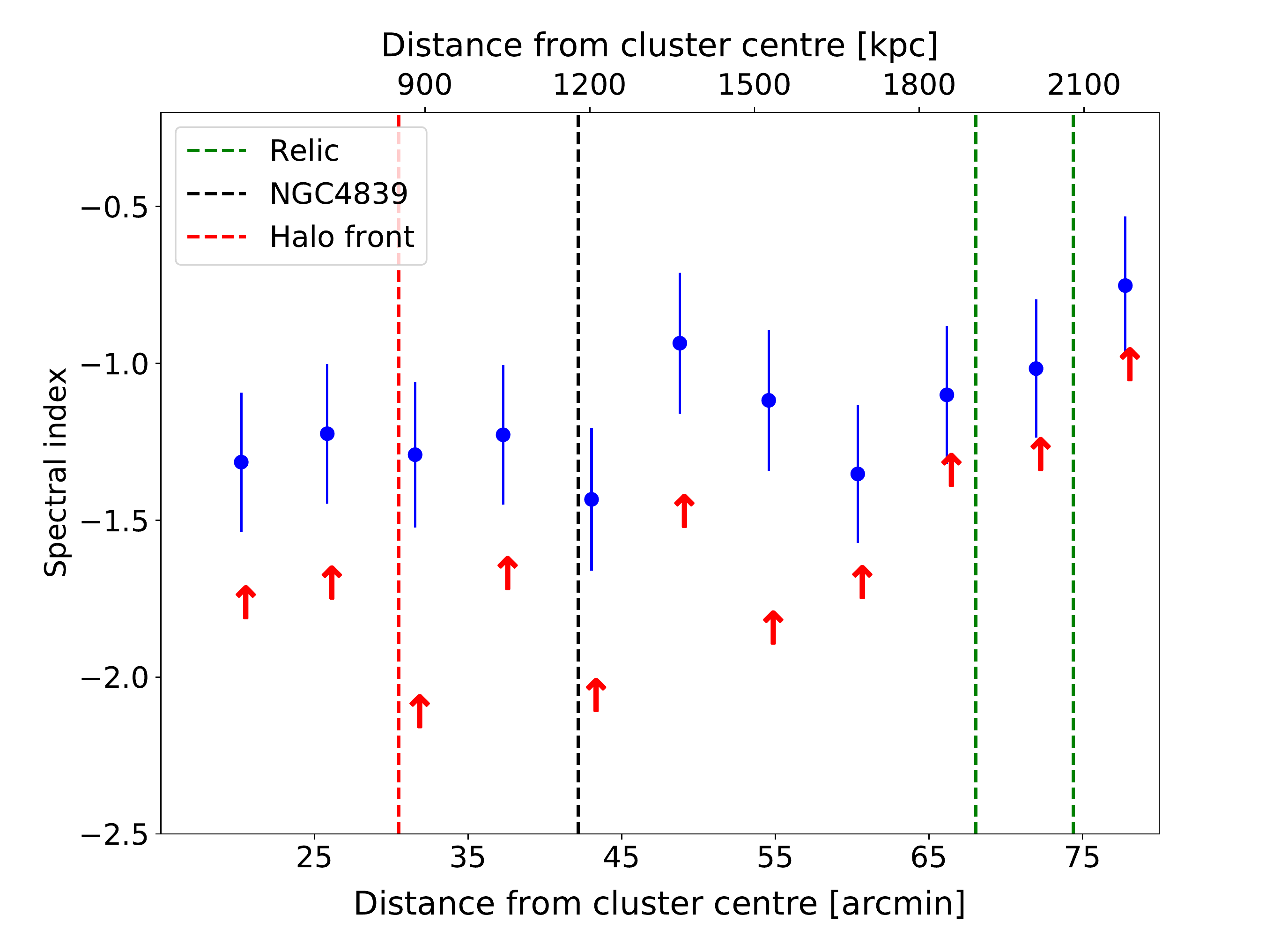}
\caption{Left panel: WSRT image at resolution of $150\arcsec \times 100\arcsec$. Contours start at 1.2 mJy/beam and are spaces by a factor 2. Boxes show the regions used to derive the average spectral index profile shown in the right panel. Right panel: spectral index profile along the bridge. Blue points refer to the emission detected at both 326 MHz and 144 MHz, red arrows show the lower limits derived using the emission detected at 144 MHz and 326 MHz above 2$\sigma_{rms}$.}
\label{fig:spix_profile}
\end{figure*}

We have used WSRT archival data \citep{Venturi90} to create new images of the Coma cluster in order to derive spectral index information in the bridge region. 
This is a fundamental quantity to understand the origin of the emission.
WSRT data have been reimaged  to achieve a resolution of 150$\arcsec \times$100$\arcsec$, that is the best compromise between resolution and sensitivity to extended emission. LOFAR data have been reimaged accordingly, matching the UV-range of WSRT observations (minimum baseline$\sim$ 40$\lambda$) and convolved to the same restoring beam. The WSRT image is shown in Fig. \ref{fig:spix_profile}, left panel, and the properties of the bridge as derived from this image are listed in Tab. \ref{tab:bridge}.\\
As mentioned above, the observations P195+27 has higher sensitivity than P192+27, and is less affected by artefacts from the subtraction of sources close to the bridge. When the images are convolved to a resolution of 150\arcsec $\times$ 100\arcsec, the quality of the two images differs even more, indicating that the short baselines of P195+27 are better calibrated. Hence, in the following analysis of the spectral index properties, we will only use the observations from the pointing P195+27.\\
For a better comparison between the two images, we have identified and masked the sources in the WSRT and LOFAR at low resolution.
 Ideally, contaminating sources should be identified in high resolution images and either masked or subtracted from the UV data. However, LOFAR and WSRT data have a very different resolutions, and the weak sources identified in the high resolution LOFAR image are blended and confused with the bridge diffuse emission once the LOFAR image is convolved to the WSRT restoring beam. One possibility would be to identify the sources in the high resolution LOFAR image, rescale their flux density and subtract it from the WSRT visibilties. We decided against this approach, as the spectral index of sources is uncertain, and not constant throughout the sources when they are unresolved. In addition, we can not exclude they would have an inverted spectrum between 144 MHz and 323 MHz. Hence, we decided to adopt a simpler approach, and identify and mask only the sources that are visible at low resolution.
The WSRT image is shown in Fig.~\ref{fig:spix_profile}. Since the brightness at 144 MHz is rather uniform (see Fig.~\ref{fig:2res}), this means that we can measure the spectral index between 144 MHz and 326 MHz only in the flattest spectral regions, and obtain limits in the remaining regions.
The average spectral index in the area where WSRT observations show emission above 2$\sigma_{\rm rms}$,
is $\alpha_{patches}=-1.3\pm0.2$. 
This value can be contaminated by the presence of point-like sources that are not perfectly masked out from the bridge emission, and that usually have a spectrum flatter than $-1.2$.\\
To provide conservative constraints to the spectrum of the bridge, we
can assume two extreme scenarios:
\begin{itemize}
    \item The emission of the bridge at 326 MHz is completely detected by WSRT observations above 2 $\sigma_{rms}$ i.e. 
    no significant further emission would be detected by deeper observations at 326 MHz.
    We can compute the average spectrum of the bridge considering 
   the emission observed at 144 MHz above 2~$\sigma_{rms}$, and the emission observed at 326 MHz above 2$\sigma_{rms}$.
     In this case, we obtain $\alpha_{L} = -1.6 \pm\ 0.2$. This value must be considered a lower limit to the real spectrum of the bridge.
\item The emission at 326 MHz falls just below 2 $\sigma_{rms}$. 
Hence, we can consider the emission at 326 MHz as the sum of two terms: $ S_{WSRT}= S_{meas} + 2 ~\sigma_{rms} ~ (N_{beam}^{L} - N_{beam}^W) $, where $S_{meas}$ is the measured flux density above 2~$\sigma_{rms}$, $N_{beam}^{L}$ indicates the bridge area detected above 2$\sigma_{rms}$ by LOFAR in beam units and similarly $N_{beam}^W$ indicates the bridge area detected above 2$\sigma_{rms}$ by WSRT in beam units.
In this case, we obtain  $\alpha_{U} = -1.4 \pm 0.2$, that is an upper limit to the spectral index of the bridge.
\end{itemize}
These two scenarios are both extreme and allow us to constrain the average spectrum of the bridge as detected by LOFAR: $-1.4 \pm0.2 \leq \alpha \leq -1.6 \pm 0.2$.
The spectral indexes derived in this section are listed in Tab. \ref{tab:alpha}.\\

In Fig.~\ref{fig:spix_profile}, the spectral index trend in the region of the bridge is shown. We can see that the spectral index steepens from the relic's outer edge towards the bridge, and does not show any particular trend within the bridge itself. In the region between the source NGC~4839 and the Coma halo, it has an average value of $-1.3 \pm 0.2$.\\
Also these values are representative of the flattest regions of the bridge, that are detected at both 144 MHz and 326 MHz. We can place a lower limit on the spectral index in each box considering the emission detected in the LOFAR image above 2$\sigma_{rms}$. These values are shown with upwards arrows in Fig.~\ref{fig:spix_profile}.

\begin{table}
\caption{$I_R - I_X$ correlation}
\begin{tabular}{lccc}
\hline\hline
Best fit $\beta$ & 25\% -- 75\% & $Corr$ & Grid size \\
0.5 & 0.6 -- 0.4 &   0.67 & 90\arcsec $\times$ 90\arcsec \\
0.4  & 0.6 -- 0.3  &   0.68 & 135\arcsec $\times$ 135\arcsec \\
\hline\hline
\multicolumn{4}{l}{Col 1: Best-fit slope; Col 2: 25th and 75th percentile of the }\\
\multicolumn{4}{l}{posterior distribution for the fit slope.}\\
\multicolumn{4}{l}{Col 3: Pearson correlation coefficient}\\
\multicolumn{4}{l}{ Col 4: Grid size used to compute $I_X$ and $I_R$.}
\end{tabular}
\label{tab:corr}
\end{table}

\begin{figure*}
\includegraphics[width=1.1\columnwidth]{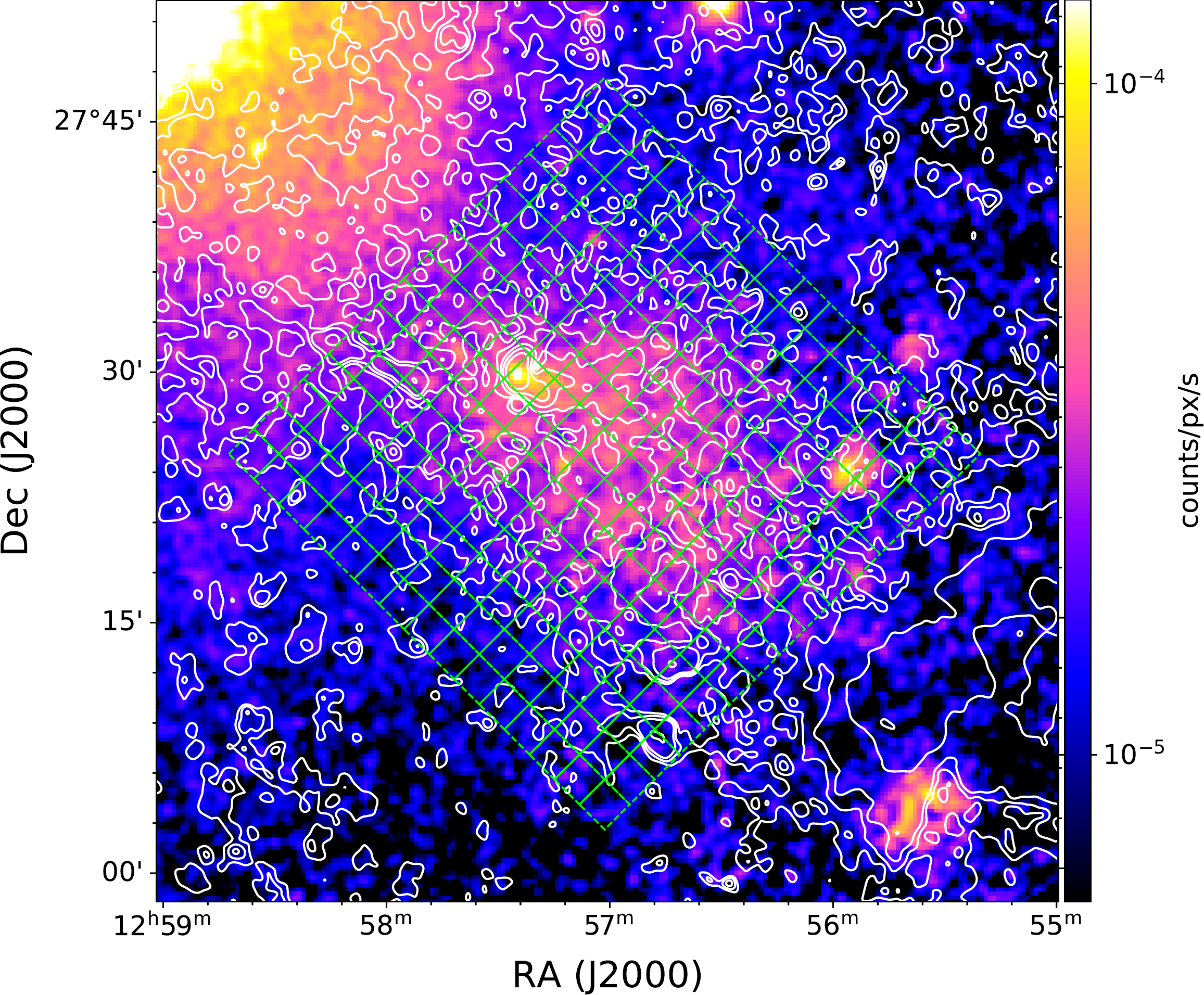}
\includegraphics[width=1\columnwidth]{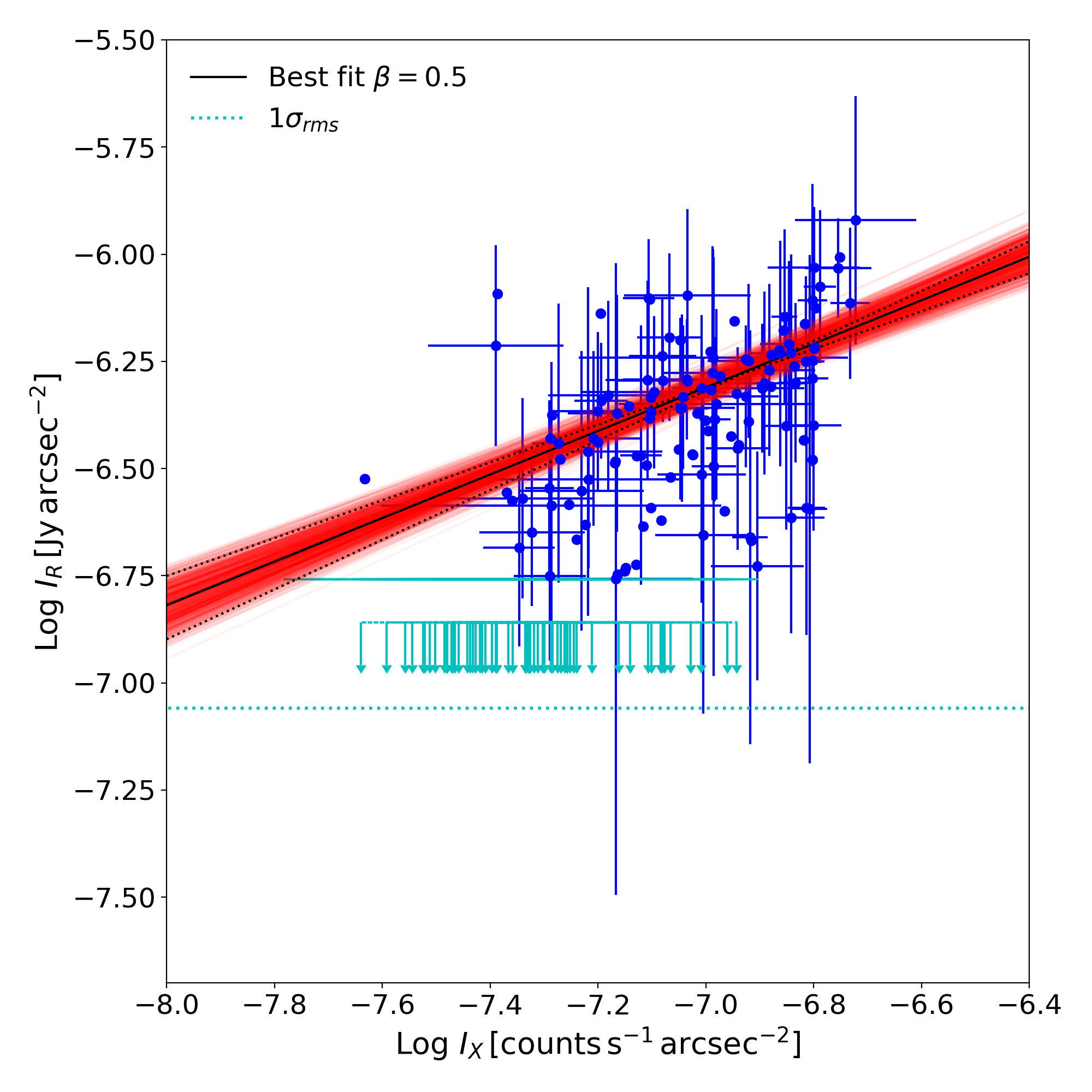}
\caption {Left panel: Colors show the X-ray emission in the Coma bridge region from ROSAT observations. Contours show the LOFAR emission from the bridge at 45\arcsec resolution, used for the analysis of the $I_R-I_X$ correlation. Contours are shown at $(3,5,9,27,82,243) \sigma_{rms}$, where $\sigma_{rms}= 0.2 \rm{mJy/beam}$. Boxes show the grid used for the analysis (90\arcsec $\times$ 90\arcsec). 
Right panel: Radio - X surface brightness correlation in the bridge. Blue lines show the statistical errors for the two quantities. For clarity, errors are shown only every two points.
The black line shows the best-fit correlation, black dotted lines show the 25th and 75th percentile of the posterior distribution. Cyan arrows are the $2\sigma_{rms}$ upper limits. The cyan dotted line marks $1\sigma_{rms}$.} 
\label{fig:corr}
\end{figure*}

\subsection{Radio and X-ray correlation}
\label{sec:correlation}
Radio and  X-ray emission have been found to be spatially correlated in some radio halos \citep[e.g.][]{Govoni01,Hoang19,Botteon20}. The existence or absence of this correlation can give important insights into the spatial distribution of the magnetic field and emitting particles with respect to the thermal gas. 
Here, we compute this correlation in the Coma bridge using LOFAR data at 144 MHz. Since we use the radio image at 45\arcsec resolution, we have smoothed the X-ray image accordingly.\\
Let $I_R$ and $I_X$ be the radio and X-ray surface brightness computed in each cell of a regular grid that we have used to analyse the bridge (see Fig.~\ref{fig:corr}). The grid is composed of a large number of cells of different sizes, as explained below and specified in Tab.~\ref{tab:corr}. Cells intercepting residual emission from AGN and radio sources have been identified and excluded from the analysis. \\
To search for a possible correlation, we have performed a fit in log-log space:
\begin{equation}
\log(I_R)= \beta \log(I_X) + \alpha_c .
\end{equation}
We have used a hierarchical Bayesian model \citep{Kelly07}, that allows us to perform linear regression of $I_R$ on $I_X$ accounting for 
intrinsic scatter in the regression relationship,  censored data, possibly correlated  measurement errors, and selection effects ( e.g. Malmquist bias). This method derives a likelihood function for the data, hence we consider the mean of the posterior distribution as the best-fit slope.
Following \cite{Botteon20}, we considered as upper limits the radio values that are below 2$\sigma_{\rm rms}$. The upper limits are properly taken into account in the fit. 
 The fit has been made with different grids, with cells of $90\arcsec \times 90\arcsec$, and $135\arcsec \times 135\arcsec$, corresponding to 7 and 16 times the beam area, respectively.\\
In Fig.~\ref{fig:corr}, the data are shown together with the best fit line, and the 25th and 75th percentile lines of the fitted posterior regression slope distribution $\beta$. In Table \ref{tab:corr}, the results of the fits are listed. $I_R$ and $I_X$ show a moderate positive correlation (Pearson correlation coefficient equal to 0.67). A positive correlation is expected if the emission originates from the same physical volume, i.e. the radio emission is as volume-filling as the X-ray emission and we are not observing radio and X-ray emission in the same region because of projection effects.\\



\begin{figure}
\includegraphics[width=1.1\columnwidth]{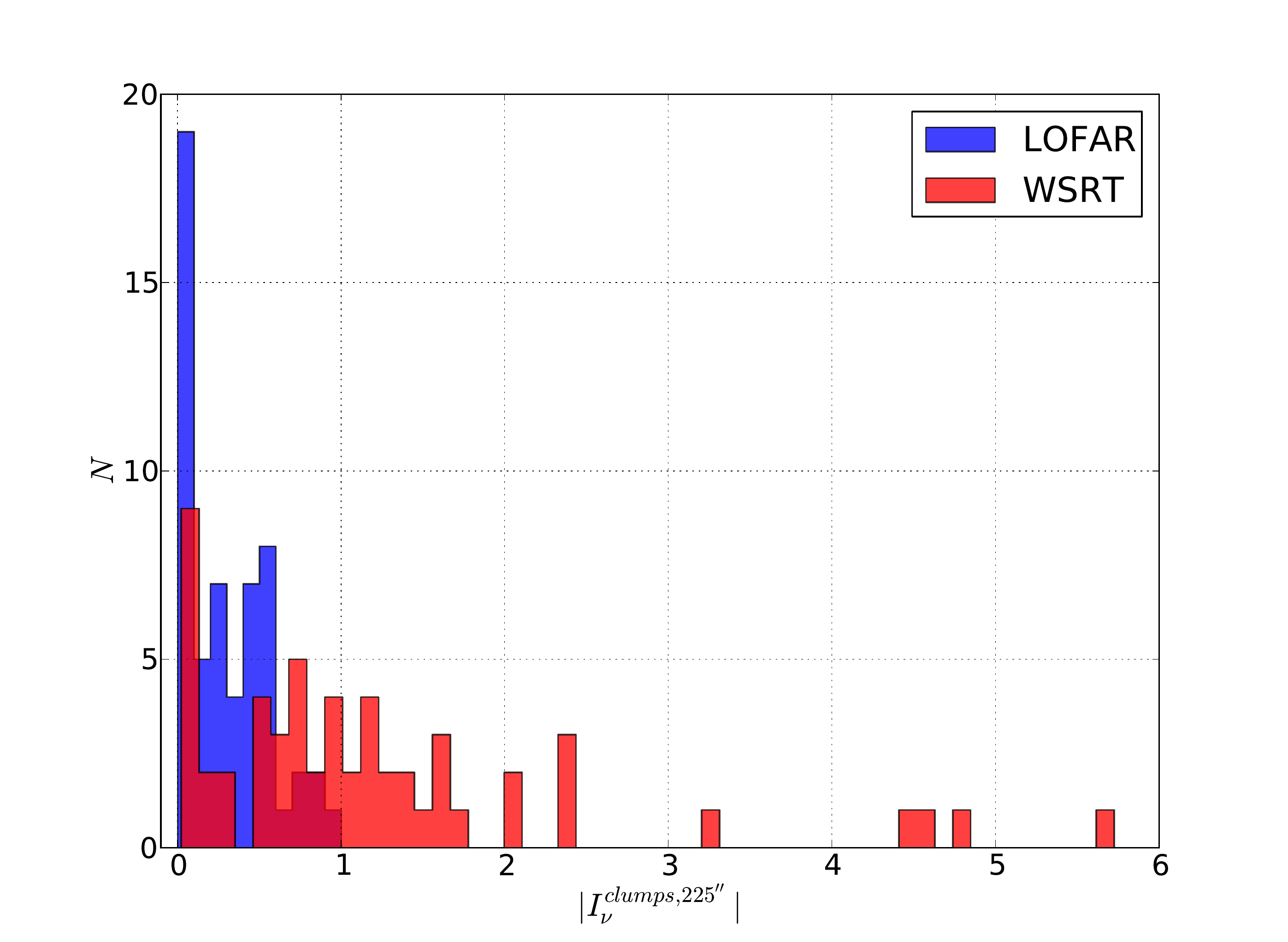}
\caption{Histogram of the quantity $ I^{clumps,225\arcsec}$ for LOFAR and WSRT observations.}
\label{fig:clumps}
\end{figure}

\subsection{Distribution of emission at different frequencies}

Turbulence mechanisms for particle acceleration predict a high-frequency cut-off of the radiating electrons \citep[e.g.][]{BrunettiLazarian16}. Specifically, super-Alfvenic solenoidal turbulence has been used to explain the emission in large-scale radio bridges, and according to this model the cut-off synchrotron frequency depends  on  the  turbulent  energy  density  and on turbulent  energy in the bridge.
Hence, particles that emit radiation at high frequencies are only those accelerated in the most turbulent regions.
It follows that the properties of the radio emission in the bridge change with frequency, and specifically  the emission at lower frequencies should be more volume-filling than the emission at higher frequencies. We can test this prediction for the case of the Coma bridge, using LOFAR and WSRT data.\\ 
To quantify the volume filling factor of the emission in the bridge, we have used the same images used for the spectral index analysis. The area emitting above 2$\sigma_{\rm rms}$ at 326 MHz is 60\% of the area above 2$\sigma_{\rm rms}$ at 144 MHz. However, this value could be affected by the different sensitivity of the two observations.\\
 To  quantify the spatial distribution of the emission at the two frequencies and evaluate whether its distribution is clumpy or smooth, we have convolved our images $I(x,y)$ with  Gaussian kernels of increasing width ($G_{X}$), where $X$ indicates the Gaussian FWHM in arcsec. We have then used a normalized unsharp-masking filter, computing:
\begin{equation}
 I^{\rm clumps,X}_{\nu}(x,y)=\frac{I(x,y) - I(x,y) * G_{X}(x,y)}{I(x,y)*G_{X}(x,y)},
\end{equation}
where $\nu$ indicates the observing frequency, and $(x,y)$ indicate the pixels in the image. $ I^{\rm clumps,X}_{\nu}(x,y)$  enhances the emission on angular scale $X$ convolved with the restoring beam of the image. 
By comparing the distribution of $I^{\rm clumps,X}_{\nu}$
at different frequencies, we can estimate the level of clumping in the radio emission as a function of frequency. 
We have computed $I^{clumps,X}_{\nu}(x,y)$ for $X=120\arcsec, 225\arcsec,315\arcsec$. For each $ I^{\rm clumps,X}_{\nu}(x,y)$, we have computed the distribution of the flux density at 326 and 144 MHz in the bridge area, limited to cells where radio emission above 2$\sigma$ was observed over a region larger than the restoring beam. Two example distributions are plotted in Fig.~\ref{fig:clumps} for $X=225\arcsec$. The distribution of WSRT data shows a long tail at high values, possibly indicating that the emission is clumpier at 326 MHz than at 144 MHz. Similar trends are obtained with 
$X=120\arcsec$ and $X=315\arcsec$. However, the different sensitivity of the two observations could play a role here, as we note that the LOFAR image at high resolution is also quite filamentary. Deeper and higher resolution observations at frequencies higher than 144 MHz would be required to derive robust conclusions from this analysis.

\section{Numerical simulations of the radio bridge}
\label{sec:sim}
Several independent works indicate that the SW sector of the Coma cluster is a dynamically active region \citep[e.g.][]{Lyskova19,Malavasi20,Mirakhor20}
where NGC~4839 is at its second infall towards the Coma cluster, likely after reaching the apocenter \citep{Lyskova19}, a shock wave is  moving towards SW from this group, powering the radio relic \citep{OgreanComa}, and streams of gas are accreting along a cosmic filament that connects Coma and A1367 \citep{Mirakhor20}.
All these motions could inject turbulence in the region of the bridge, especially at the West of the NGC~4839 group, where radio emission is observed.
 We note that the emission observed between the group and the halo has a lower surface brightness than the emission between the group and the relic (see Fig. \ref{fig:profile}), and we cannot even exclude that the bridge is not connected to the halo (see Sec.\ref{sec:bridge}). This could be explained considering the motion of the group in its orbit around the cluster. During the first passage towards the apocenter, the motion of the group could have triggered particle acceleration and originate the emission that we now detect with a low surface brightness between the group and the halo. While the second passage of the group in the bridge region, after reaching the apocenter, could have originated the brighter emission that we observe at the West of the group. \\
We can derive a rough estimate of the turbulent energy flux and compare it to the non-thermal luminosity. 
This comparison allows one to understand if the observed radio luminosity can be powered by the damping of the turbulent energy flux into relativistic particles. The turbulent energy flux in the volume $V$,  $ V \times F$,  is given by:
\begin{equation}
V \times    F= V \cdot \frac{\rho c_s^3 \mathcal M^3}{2L}
\end{equation}
where $V$ is the bridge volume, $\rho$ is the gas density, $c_s$ is the sound speed, $\mathcal M$ is the Mach number, and $L$ is the turbulence injection scale. Assuming that the group is the main cause for the injection of turbulence, we can consider $L \sim \rm{300 kpc}$. We can also estimate $c_s \sim 1500 $ km/s (from an average temperature of 8 keV, \citealt{Simionescu13} in the SW sector of Coma). The volume of the bridge can be approximated by  a cylinder with a circular base of 300 kpc radius and 900 kpc height, and the gas density can be roughly approximated by $n_e \sim 5\cdot 10^{-4} \rm{cm^{-3}}$ \citep{Simionescu13}\footnote{
since the asymmetric geometry complicates the deprojection along the bridge, we have assumed that the density here can be estimated scaling the deprojected density along other relaxed directions by the square root of the ratio between the X-ray brightness of the corresponding regions.}.\\ 
A fraction, $\eta$, of the energy flux $F$ can be dissipated into the acceleration of mildly relativistic electrons and eventually into non-thermal emission (i.e. synchrotron and Inverse Compton luminosity):
\begin{equation}
    \eta F V \sim L_{tot} \sim L_{S}(1 + (B_{cmb}/B)^2)  
\end{equation}
where $L_S\sim ~1.4\cdot10^{40} ~\rm{ergs/s}$, considering a power-law power spectrum with $\alpha=-1.5$ in the range 20 MHz - 2 GHz, and $B_{cmb} = 3.25 (1+z)^2 ~\rm{\mu G }$.
This allows one to constrain the Mach number of turbulence required for a given efficiency $\eta$:
\begin{equation}
    \mathcal M \eta^{1/3} \sim  0.01 \left( 1+ 0.25 \cdot \frac{B_{cmb}^2}{ B_2^2} \right)^{1/3} L_{0.3 } ^{1/3} V_{1} \rho_{5\cdot10^{-28}} ^{-1/3}  c_{s_{1500}}^{-1}
\end{equation}
Here $B_2$ is the magnetic field normalized to 2 $\mu$G, as derived from Faraday RM studies \citep{Bonafede13}, $L_{0.3}= L/0.3 ~\rm{Mpc}$, $V_{1}$ is the bridge volume normalised to 1 $\rm{Mpc}^3$, $\rho_{5\cdot10^{-28}}$ is the gas density normalised to $(5\cdot10^{-28}~\rm{ g/cm^3)}$, and $c_{s_{1500}}$ is the sound speed  normalised to 1500 ${\rm km/s}$.
Assuming $\eta \lesssim 10^{-4}$, as usually required to explain radio halos \citep{BrunettiLazarian07}, 
we derive that $\mathcal M \gtrsim 0.3$ is enough to power the radio bridge.
As $\eta \sim 10^{-4}$ corresponds to a turbulent energy budget that is $\sim$ 5\% of the thermal energy budget, we conclude that the available energy budget is in principle enough to explain the radio emission.\\

The exact mechanism through with turbulent motions accelerate particles is yet unconstrained. A viable channel of acceleration is the Transit Time Damping (TTD), mediated by compressive turbulent modes \citep[e.g.][]{BrunettiLazarian07}, often assumed to explain radio halos. 
 However, cosmological simulations suggest that the largest budget of turbulent kinetic energy in the ICM  is solenoidal \citep{Miniati14,Vazza17}, with compressive modes contributing  up to 30$\%$ only at $r \geq 0.5 ~r_{vir}$.
Cosmological simulations also show that solenoidal turbulence is dissipated in the ICM at a much smaller rate than compressible turbulence, because of its flatter velocity spectrum, providing a possible energy reservoir for particle acceleration in the ICM \citep[][and ref tehrein]{Vazza17}. Solenoidal turbulence can contribute to particle acceleration in the ICM.  \citet{BrunettiLazarian16} have proposed a model in which particles are re-accelerated stochastically diffusing across regions of magnetic reconnection and dynamo, which is a viable mechanism for large-scale super-Alfv\'enic solenoidal turbulence.
Both TTD and acceleration by solenoidal turbulence may be viable mechanisms for particle re-acceleration in the ICM, and up to now, it is not clear which of the two mechanisms dominates in the ICM. Their relative importance depends on the ratio of compressive and solenoidal motions that are available in the ICM and on ICM microphysics (see \citealt{BrunettiLazarian16} and \citealt{BV20}for a comparison between the two mechanisms.)
Recently,  \citet{BV20} have shown that re-accelreation mediated by solenoidal turbulence, also called Adiabatic Stochastic Acceleration, may be responsible for the steep spectrum and volume filling synchrotron emission observed in bridges in-between merging galaxy clusters \citep[][]{Govoni19}.  In the same work, it is shown that Adiabatic Stochastic Acceleration is more efficient than TTD, provided that turbulence is subsonic, $\mathcal{M} \leq 0.5$ \citep[][]{BV20}.
In the following, we will investigate under which conditions this mechanism could work in the Coma bridge.

\subsection{Cosmological simulation}
The cosmological simulation that we will use in the following analysis is taken from 
the suite of magneto-hydrodynamical (MHD) non-radiative high-resolution simulations presented in \citet{va18mhd} and \citet{dom19}.  The simulations have been made with ENZO \citep[][]{enzo14} and include eight levels of Adaptive Mesh Refinement (AMR) to  increase the spatial and force resolution in a large fraction of the simulated cluster volume, up to $\Delta x=3.95 \rm ~ kpc/cell$. One of the simulated clusters has properties similar to Coma  \citep[see][for further details]{va18mhd}. In particular, the simulation evolves a uniform primordial magnetic field of $B_0=10^{-10} \rm ~G$ (comoving) and the resulting Faraday Rotation Measures (RM) profile of the cluster is in agreement with the one derived from observations \citep{Bonafede10,va18mhd}.

Hence, this Coma-like cluster is suitable for comparison with LOFAR observations and will be used in the calculations below.\\
We focus on a region at the periphery of the simulated cluster, where matter is accreted onto the main cluster, and where it has been found that two satellites have an additional velocity component perpendicular to the cluster radius (see Fig. \ref{fig:sim}).
Hence, even if the specific dynamical status of the simulated cluster is not exactly the same as the observed bridge,
the simulation represents a realistic environment to investigate the origin of the radio emission. In the following, we refer to the region in the simulation where matter is accreting onto the main cluster as to the simulated bridge.\\
In order to study the plasma conditions that could lead to the (re)acceleration of radio emitting electrons in the bridge, we  separated turbulent fluctuations, $\delta V$, from bulk motions on larger scales, using an iterative three-dimensional filtering technique. Specifically, we extracted solenoidal, $\nabla \cdot \vec{v}=0$, and compressive, $\nabla \times \vec{v}=0$, turbulent components using the Hodge-Helmholtz projection in Fourier space \citep[e.g.,][]{va17turb}. This allows us to compute the local turbulent energy flux, $F$ due to solenoidal and compressive modes (e.g. \citealt{2020MNRAS.495..864A}).

\subsection{Turbulent re-acceleration in the bridge}
\label{sec:acceleration}

\begin{figure}
\includegraphics[width=1.0\columnwidth]{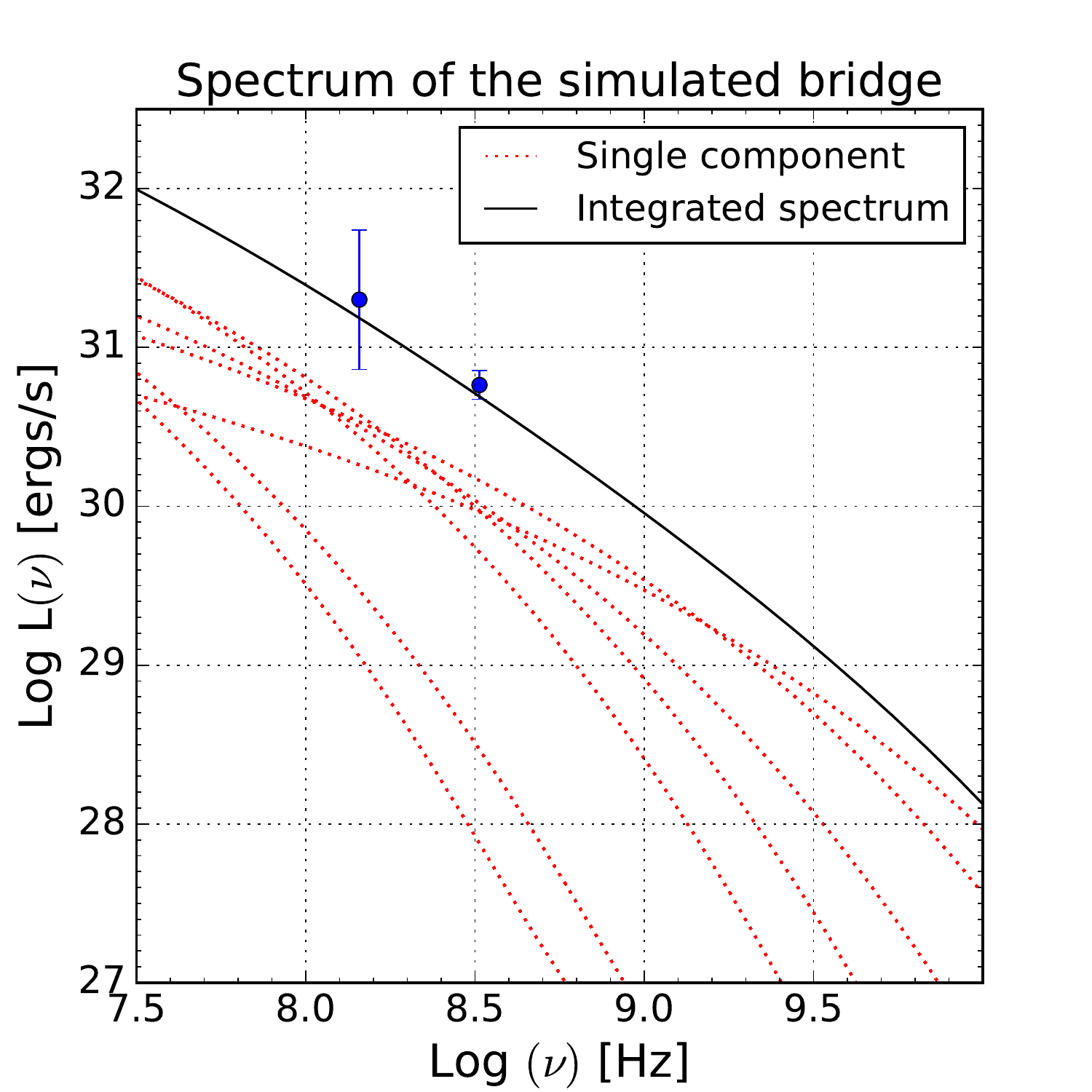}
\caption{Spectrum of the radio emission in the bridge (solid black line) obtained assuming a single-zone model and evolving Fokker-Planck equations for a population of seed electrons (see Sec. \ref{sec:acceleration} for details). Dotted red lines show the spectrum of single components, computed with different acceleration times. The black line is obtained summing the contributions of the single components. The blue dots mark the observed monochromatic power at 144 and 326 MHz.}
\label{fig:spectrum_th}
\end{figure}

Here, we compute the electron acceleration rate due to turbulence using the cosmological simulations presented above.
In principle, the evolution of radio-emitting particles  should be computed with a Lagrangian approach \citep[e.g.][]{Wittor17b}, and then integrating in time their spectral energy evolution using Fokker-Planck equations \citep[e.g.][]{2014MNRAS.443.3564D}.
However, this approach is numerically challenging and beyond the scope of this paper. Hence, we adopt a simplified approach to estimate the radio emission as a function of frequency from the simulated bridge.
Following \cite{BV20}, 
we have measured the distribution function of acceleration times in boxes of $(500 ~\rm kpc)^3$ , i.e. $30^3$ cells, in the simulated bridge. The acceleration time is
\begin{equation}
    \tau_{\rm acc}= 1.25 \cdot 10^5 (\rm Myr) 
    {{ L_{0.5} B^*_{\mu G} }\over{ \sqrt{n_{th3}} \delta V_7^3}},
    \label{eq7}
\end{equation}




\noindent
where $n_{\rm th3}$ is the thermal particle number density normalised by $10^{-3}\rm cm^3$, $\delta V_7$ is the turbulent velocity (in units of $10^7 \rm ~cm/s$) measured on a scale $L_{0.5}$ in each cell, and $B^*_{\mu G}$ is the magnetic field strength in $\rm \mu G$.
Following \citet{BV20}, we account for the effect of unresolved dynamo amplification in the simulation by setting in every cell: $B^*= {\rm max} \{ B_{\rm sim}, \sqrt{4 \pi \rho \eta_B \delta V^2\}}$, where $B_{\rm sim}$ is the magnetic field measured in the simulation, and $\eta_B$ is the fraction of turbulent kinetic energy flux that is converted into magnetic 
fields. As discussed in \citet{BV20}, we have assumed $\eta_B=0.03$. \\
Then, we have sampled the distribution function of acceleration times 
with a step of $\Delta \log(\tau_{\rm acc})=0.2$, obtaining 10 values for $\log(\tau_{\rm acc})$.
For each $\tau_{\rm acc}$, we have evolved an initial spectrum of relativistic electrons solving the Fokker-Planck equations for a turbulent eddy-turnover time and assuming a single-zone model, i.e. using  the average thermal densities and magnetic field in the simulated box.
We have considered the combined effect of acceleration, radiative and Coulomb losses.\\
Finally, we have computed the integrated synchrotron spectrum combining the contributions from the spectra derived from different acceleration times and assuming a single value for the magnetic field ($ \langle B^* \rangle =0.5 \mu\rm{G}$). These spectra have been weighted according to the distribution functions of acceleration times, and are shown in Fig. \ref{fig:spectrum_th}\footnote{Only seven components are visible in the plot, as the remaining three have a luminosity outside the range shown in the figure and their contribution to the integrated spectrum is negligible.}. \\
The synchrotron luminosity produced by a region of 1 Mpc$^3$ is shown in Fig.~\ref{fig:spectrum_th}. We have to assume that the fossil population of electrons have an initial energy budget that is $3 \cdot10^{-4}$ with respect to the thermal gas in order to match the observed luminosity\footnote{The exact shape of the 
initial spectrum does not affect significantly the final result.}.
This value is almost a factor 10 higher than the one used \citet{BV20} for the bridge in A399-A401.\\
This discrepancy can be due to several reasons, such as the magnetic field properties in the bridge, as the simulation is only a rough approximation of the physical conditions in the Coma bridge. Another possibility is that the model proposed by \citet{BV20} can only explain a fraction of the observed emission. 
More interestingly, this discrepancy could suggest that a key role is played by the large number of seed electrons released in the medium, either by the NGC~4839 tail during its motion and by other radiogalaxies present in the bridge region \citep[see e.g.][]{Venturi90,Giovannini91}. In this case, we can conclude that the Coma bridge is radio bright thanks to a combination of turbulent motions and to the presence of tailed radiogalaxies that populate the medium with a high density of mildly relativistic electrons.\\
 The spectrum of the radio Luminosity between 150 MHz and 330 MHz is $\alpha= -1.5$, within the range allowed by our observational constraints.

\subsection{Simulated radio emission from the bridge}

Using some further assumptions, we can investigate whether the properties of the bridge resulting from turbulent re-acceleration would be in agreement with observations. We have obtained the synchrotron emissivity from each cell along the simulated bridge, assuming that a fraction of the turbulent energy flux is converted into particle 
energy flux, and then into synchrotron, and Inverse Compton radiation:

\begin{equation}
    \epsilon (\nu)  \propto n_{\rm th3} {{\delta V_7^3 }\over{ 
    L_{0.5} }} {{ ( B^*_{\mu G} )^2}\over
    {( B^*_{\mu G} )^2 + [3.25 (1+z)^2]^2}}
    \Theta(\nu_s) .
    \label{eq11}
\end{equation}

\noindent
Here, $\nu_s$ is the maximal frequency for which synchrotron radiation is emitted in each cell. 
We have put $\Theta(\nu_s)=1$ for $\nu_s > \nu_0$ and $=0$ otherwise, in order to select 
only the cells that contribute to the synchrotron radiation observed 
at the frequency $\nu_0$.\\
Following \citet{Cassano10}, we have assumed $\nu_s \sim 7 \nu_c$, $\nu_c$ being the critical synchrotron frequency, $\nu_c \sim 4.6 \gamma_{max}^2 B^*_{\mu G} \rm ~Hz$, where $\gamma_{max}$ is the Lorentz factor
of electrons when radiative losses balance the acceleration rate.
Hence, the critical frequency is:
\begin{equation}
    \nu_{c} (\rm MHz) = 1.4 \cdot10^{-3} {{
    \Phi_7 n_{th3} \delta V_7^6}\over{
L_{05}^2 B^*_{\mu G} \beta_{B,z}^{2} }}
    \left( 1 - {{\tau_{acc}}\over{\tau_{adv}}}
    \right)^2 ,
    \label{eq10}
\end{equation}
where $\beta_{B,z}$ is defined as:
\begin{equation}
    \beta_{B,z}= ( {{B^*_{\mu G} }\over{3.25}} )^2 \times (1+z)^4, 
\end{equation}
and  $\tau_{\rm adv}$ is the expansion/compression timescales, defined as
\begin{equation}
    \tau_{\rm adv} =
    7720\, {\rm Myr}\left (8.11 \cdot 10^{16} 
    \nabla \cdot \vec{v} \right )^{-1} ,
    \label{eq4}
\end{equation}
\noindent where $\vec{v}$ is the velocity field measured in $\rm km/s$, 
and $\Phi_7 \sim 0.71-1.14$. We refer the reader to \citet{BV20} and \citet{BrunettiLazarian16} for further details.

\begin{figure}
\includegraphics[width=1.0\columnwidth]{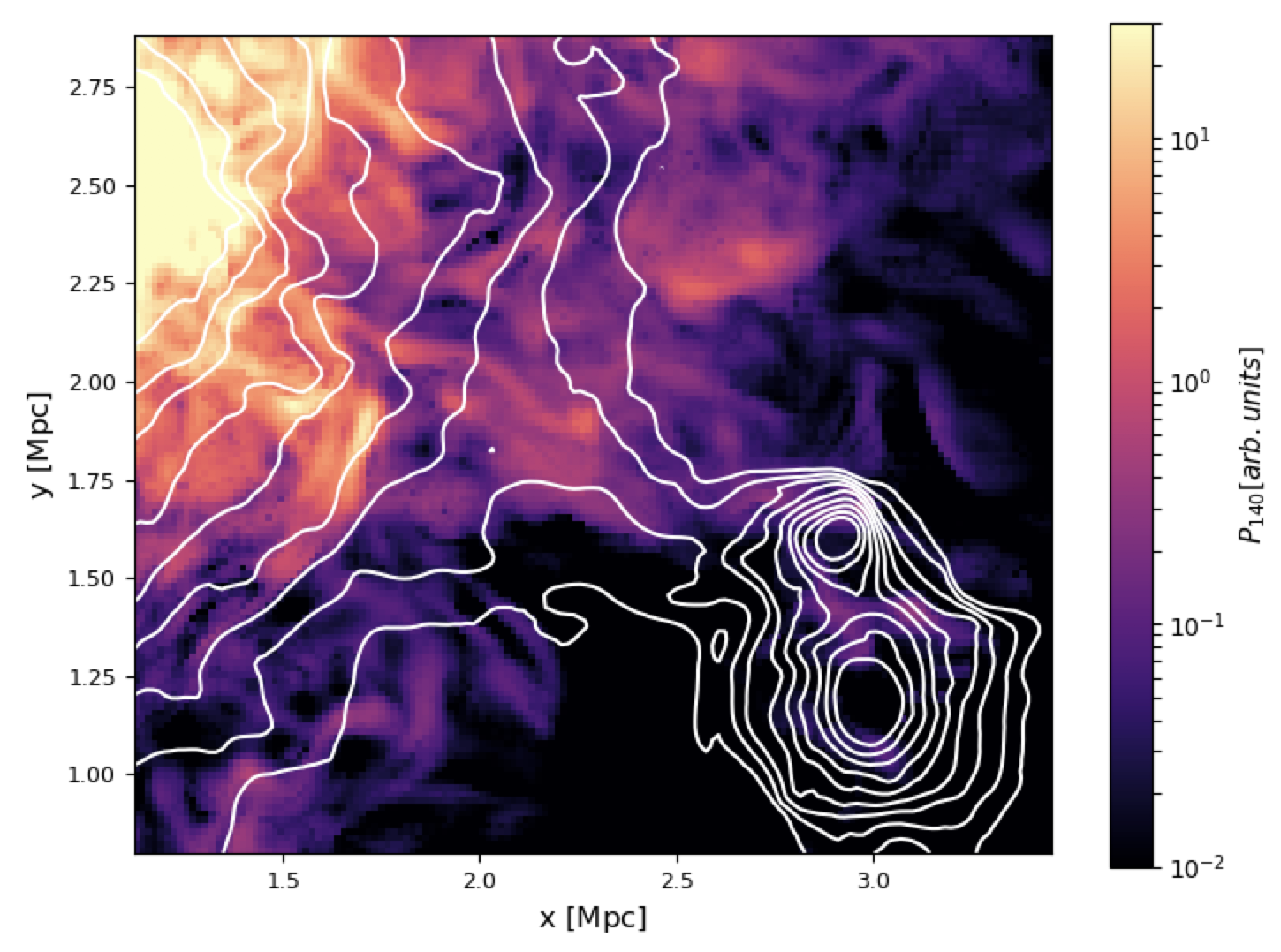}
\caption{Map of the simulated radio emission at 140 MHz (colors) and simulated X-ray emission in the [0.5-2.0] keV range (contours) for a Coma-like galaxy cluster, based on the scenario described in Sec. \ref{sec:sim} }
\label{fig:sim}
\end{figure}

Fig.~\ref{fig:sim} shows the simulated radio emission derived from the above formula, for  $\nu=140 \rm ~MHz$, overlayed onto the X-ray emission contours in the [0.5-2.0] keV energy range (obtained assuming a uniform $0.3 Z_{\odot}$ metallicity and using the B-APEC model). The map zooms into a $(2 \rm ~Mpc)^2$ region in the cluster periphery, where also the profile of Faraday Rotation is compatible with the trend derived by \citet{Bonafede13} for the SW sector of Coma (the simulated bridge, see \citealt{va18mhd}). 
The simulated radio emission shows an excess spanning $\sim 2$ Mpc along the simulated bridge, with a surface brightness $\sim 10-10^2$ lower than what the same model predicts for the central radio halo.\\ 
We find that the parts of the bridge that are most suitable for the turbulent acceleration mechanism have a thermal gas density in the range $n \sim 1-3 \cdot 10^{-28} \rm ~g/cm^3$, a gas temperature of $T \sim 4-7 ~\rm keV$, and a  (solenoidal) turbulent velocity dispersion $\sigma_v \sim 500-700 ~\rm km/s$.
The gas density and temperature are in general agreement, though slightly lower, than the estimates from X-ray analysis (see \citealt{Simionescu13}, and Sec. \ref{sec:bridge}).\\
For patches with lower densities or velocities, the model predicts $\nu_s < \nu_0$. Denser substructures in this gaseous bridge produce more emission with a relatively steep spectrum ($\alpha \approx -1.2 - -1.3$). However, such self-gravitating halos might in reality turn into galaxies, causing us to overestimate their gas content.\\
Finally, we have derived from the simulation the
$I_X - I_R$ correlation, using a region comparable to the one used in Fig.~\ref{fig:corr} and the same method described in Sec.~\ref{sec:bridge}.
We find a moderate correlation in simulated quantities (Pearson correlation coefficient $=0.5$), with a sublinear scaling of $\beta=0.80$. The 5th and 95th percentile of the posterior distribution of $\beta$ are 0.82 and 0.73, respectively, which is larger than the observed value ($\sim$ 0.5, see Sec. \ref{sec:correlation}).
Even though the moderate correlation and the sub-linear scaling found in simulations are in rough agreement with the observed ones, our simulation neglects important physical details: we have considered a single-zon model, assumed that the amplification of magnetic fields via (unresolved) small-scale dynamo happens instantaneously, and that relativistic electrons are not subject to spatial diffusion on scales larger than the cell size. All these assumptions are clearly violated in the real ICM, and only with more sophisticated approaches (e.g. using tracer particles, \citealt{Donnert13}) a quantitative three-dimensional model can be obtained.

\section{Conclusions}
\label{sec:conclusions}
We have presented LOFAR HBA observations of the Coma cluster at 144 MHz. These observations allow us to study the morphology and spectral properties of the bridge of radio emission that  connects the halo to the relic. Our main conclusions are:\\
\begin{itemize}
    \item The Coma bridge is clearly detected at 144 MHz at 30\arcsec resolution, with unprecedented sensitivity and detail. The LLS of the radio bridge is $\sim$ 900 kpc, and its brightness is fairly uniform along the bridge, apart from an apparent increase when it crosses the NGC~4839 group. At 30\arcsec resolution, a gap is observed between the bridge and the halo. 
    \item Using WSRT observations, we have analysed the spectral properties of the bridge. No trend is detected along the bridge. The emission detected in the LOFAR image is only partially detected in the WSRT image. This is due to a combination of spectral properties and different sensitivity of the two observations. We can constrain the spectral index $\alpha$ to lie in the range $-1.4 \pm 0.2 < \alpha < -1.6 \pm 0.2$.
    \item Using ROSAT observations, we have analysed the correlation between the radio and X-ray surface brightnesses, finding that the two quantities are moderately correlated (Pearson correlation coefficient = 0.67), with a sub-linear scaling of 0.5, flatter than what is reported in the literature for the Coma halo. The moderate correlation might indicate that the radio emission is as volume filling as the X-ray emission.
    \item The distributions of the radio emission at 326 MHz and 144 MHz has been analysed applying to both images a normalised unsharp-mask filter. The emission at 326 MHz appears to be more clumpy with respect to the emission at 144 MHz. However, the different sensitivity of the two observations do not permit us to derive frm conclusions.
     \item{As the electron cooling length is only $\sim$200 kpc at 144 MHz, only a part of the bridge can be explained by particles that are cooling downstream from the shock that formed the radio relic at the SW of Coma. Additional re-acceleration mechanisms are necessary to explain the radio emission of the bridge.  }
    \item We have  investigated whether turbulent acceleration can be responsible for the emission in the radio bridge. 
    We find that $\eta^{1/3} \mathcal{M} \sim 0.01$ is required to match the observed radio luminosity in these models, $\eta$ being the fraction of turbulent energy flux that is damped into the acceleration of relativistic electrons.
     We have used a cosmological simulation of a Coma-like cluster to further investigate this scenario. The simulation is used to separate solenoidal and compressive turbulent components, and to derive average values for the gas density and magnetic field. Then, we have adopted a single-zone model and evolved a population of seed CR-electrons over one eddy turn-over time using Fokker-Planck equations. 
   Assuming that the initial energy ratio of the seed particles is $3 \times 10^{-4}$ with respect to the thermal gas, and making further approximations for the magnetic field amplification, we have obtained a spectral index $\alpha=-1.5$, in general agreement with observational constraints. 
   \item a ratio of $3 \times 10^{-4}$ between non-thermal and thermal particles is higher than required to explain the radio emission in other cluster environments. It could indicate that the Coma bridge is radio emitting thanks to the contribution of radio galaxies, such as the NGC~4839 radio tail, that is injecting a considerable amount of seed particles in the medium.
    \item Using the simulation of the Coma-like cluster, we have computed the synchrotron emissivity that would originate from the simulated bridge. The simulated emission shows an excess over $\sim$ 2 Mpc along the region where matter is accreting towards the cluster,  resembling observations of the Coma bridge, although a detailed comparison of the two would require a more detailed modelling of the particle energy spectrum and its evolution over time.
\end{itemize}
We conclude that turbulent reacceleration mechanisms could explain the radio bridge in the Coma cluster, where radiogalaxies as NGC~4839 are populating the medium with mildly relativistic electrons. However, some 
quantities are still poorly constrained from observations, such as the spectral index and the morphology of the emission.
Future very sensitive observations, either at lower and higher frequencies than those presented in this work, will be useful to better characterise the morphology of this bridge and its properties as a function of frequency. 

\section*{Acknowledgments}

Large part of this work has been written during the lockdown that several countries experienced in 2020 to fight the COVID-19 pandemic. We thank all the people who during these challenges times continued working in hospitals, Pharmacies, grocery stores, transports, and other essential services, allowing us to stay safe at home wondering about the mysteries of the Universe.\\
A. Bonafede, E. Bonnassieux, M.Brienza, and C. Stuardi acknowledge support from the ERC through the grant ERC-Stg DRANOEL n. 714245. A. Bonafede acknowledges support from the MIUR grant FARE SMS. 
 F. Vazza acknowledges financial support from the ERC  Starting Grant "MAGCOW", no. 714196.  
 G. Brunetti, R. Cassano, F. Gastaldello, and M. Rossetti acknowledge support from INAF mainstream project “Galaxy Clusters Science with LOFAR” 1.05.01.86.05.
 The cosmological simulations in this work were performed using the ENZO code (http://enzo-project.org). The authors gratefully acknowledge the Gauss Centre for Supercomputing e.V. (www.gauss-centre.eu) for supporting this project by providing computing time through the John von Neumann Institute for Computing (NIC) on the GCS Supercomputer JUWELS at J\"ulich Supercomputing Centre (JSC), under projects {\it stressicm} (PI F.Vazza).
A. Simionescu is supported by the Women In Science Excel (WISE) programme of the Netherlands Organisation for Scientific Research (NWO), and acknowledges the World Premier Research Center Initiative (WPI) and the Kavli IPMU for the continued hospitality. SRON Netherlands Institute for Space Research is supported financially by NWO.
A.Drabent acknowledges support by the BMBF Verbundforschung under grant 05A17STA. The J\"lich LOFAR Long Term Archive and the German LOFAR network are both coordinated and operated by the J\"ulich Supercomputing Centre (JSC), and computing resources on the supercomputer JUWELS at JSC were provided by the Gauss Centre for Supercomputing e.V. (grant CHTB00) through the John von Neumann Institute for Computing (NIC). 
A. Botteon acknowledges support from the VIDI research programme with project number 639.042.729, which is financed by the Netherlands Organisation for Scientific Research (NWO). R.J. van Weeren and G. di Gennaro acknowledges support from the ERC Starting Grant ClusterWeb 804208. 
LOFAR(vanHaarlem et al. 2013) is the Low Frequency Array designed and constructed by ASTRON. It has observing, data processing, and data storage facilities in several countries, which are owned by various parties (each with their own funding sources), and that are collectively operated by the ILT foundation under a joint scientific policy. The ILT resources have benefited from the following recent major funding sources: CNRS-INSU, Observatoire de Paris and Universit\'e d'Orl\'eans, France; BMBF, MIWF-NRW, MPG, Germany; Science Foundation Ireland (SFI), Department of Business, Enterprise and Innovation (DBEI), Ireland; NWO, The Netherlands; The Science and Technology Facilities Council, UK; Ministry of Science and Higher Education, Poland; The Istituto Nazionale di Astrofisica (INAF), Italy. This research made use of the Dutch national e-infrastructure with support of the SURF Cooperative (e-infra 180169) and the LOFAR e-infra group. The J\"ulich LOFAR Long Term Archive and the GermanLOFAR network are both coordinated and operated by the Jülich Supercomputing Centre (JSC), and computing resources on the supercomputer JUWELS at JSC were provided by the Gauss Centre for Supercomputinge.V. (grant CHTB00) through the John von Neumann Institute for Computing (NIC). This research made use of the University of Hertfordshirehigh-performance computing facility and the LOFAR-UK computing facility located at the University of Hertfordshire and supported by STFC [ST/P000096/1], and of the Italian LOFAR IT computing infrastructure supported and operated by INAF, and by the Physics Department of Turin university (under an agreement with Consorzio Interuniversitario per la Fisica Spaziale) at the C3S Supercomputing Centre, Italy.

\bibliographystyle{aasjournal}
\bibliography{master}



\end{document}